\documentclass[preprint,letterpaper,amsmath,superscriptaddress,floatfix]{revtex4}
\usepackage{amssymb,amsfonts,amsmath}
\usepackage{graphicx}
\usepackage{tabularx}
\begin{document}  %% Titlepage
%%%%%%%%%%%%%%%%%%%%%%%%%%%%%%%%%%%%%%%%%%%%%%%%%%%%%%%%%%%%%%%%%%%%%%%%%%%%%%%

\title{Lithium Subhydrides Under Pressure and their Superatom--Like Building Blocks}
\author{James Hooper}
\author{Eva Zurek}\email{ezurek@buffalo.edu}
\affiliation{Department of Chemistry, State University of New York at Buffalo, Buffalo, NY 14260-3000, USA}
\begin{abstract}
Evolutionary structure searches are used to predict a new class of compounds in the lithium--rich region of the lithium/hydrogen phase diagram under pressure. First principles computations show that Li$_m$H, $4<m<9$, are stabilized with respect to LiH and Li between 50-100~GPa. The building block of all of the lithium subhydrides is an Li$_8$H cluster, which can be thought of as a superalkali. The geometries and electronic structures of these phases is analogous to that of the well--known alkali metal suboxides. 
\end{abstract}
%\pacs{71.20Dg, 62.50.-p, 63.20.dk}% PACS, the Physics and Astronomy Classification Scheme.
%Electron density of states and band structure of crystalline solids
%http://www.aip.org/pacs/pacs2010/individuals/pacs2010_regular_edition/reg70.htm#71
%71.20.Ps	Other inorganic compounds
%%71.18.+y	Fermi surface: calculations and measurements; effective mass, g factor
%\keywords{high pressure, alkali hydrides, metallization, first principles}
\maketitle

Considerable attention has been given towards the pressure--induced synthesis of hydrogen--rich systems with unusual stoichiometries due to  their potential import in energy applications, and because they may provide routes towards hydrogen's metallization and concomitant superconductivity.\cite{Klug:2011a} A number of intriguing systems which exhibit strong intermolecular interactions under pressure, including the van der Waals solids Xe-H$_2$,\cite{Somayazulu:2010a} CH$_4$-H$_2$,\cite{Somayazulu:1996a} SiH$_4$-H$_2$,\cite{Strobel, Wang:2009a} and H$_2$S-H$_2$,\cite{Strobel:2011a} have been prepared. Pressure--induced hydrogen uptake has been observed in metals with low hydrogen solubility yielding  RhH$_2$,\cite{Li:2011a} Re$_2$H,\cite{Scheler:2011b} PtH,\cite{Kim:2011a,Zhou:2011a,Scheler:2011a} or noble metal hydrides.\cite{Ma:2012a} Yet, not much work has been devoted to studies of compounds which are hydrogen poor.

\begin{figure}
\begin{center}
\includegraphics[width=\columnwidth]{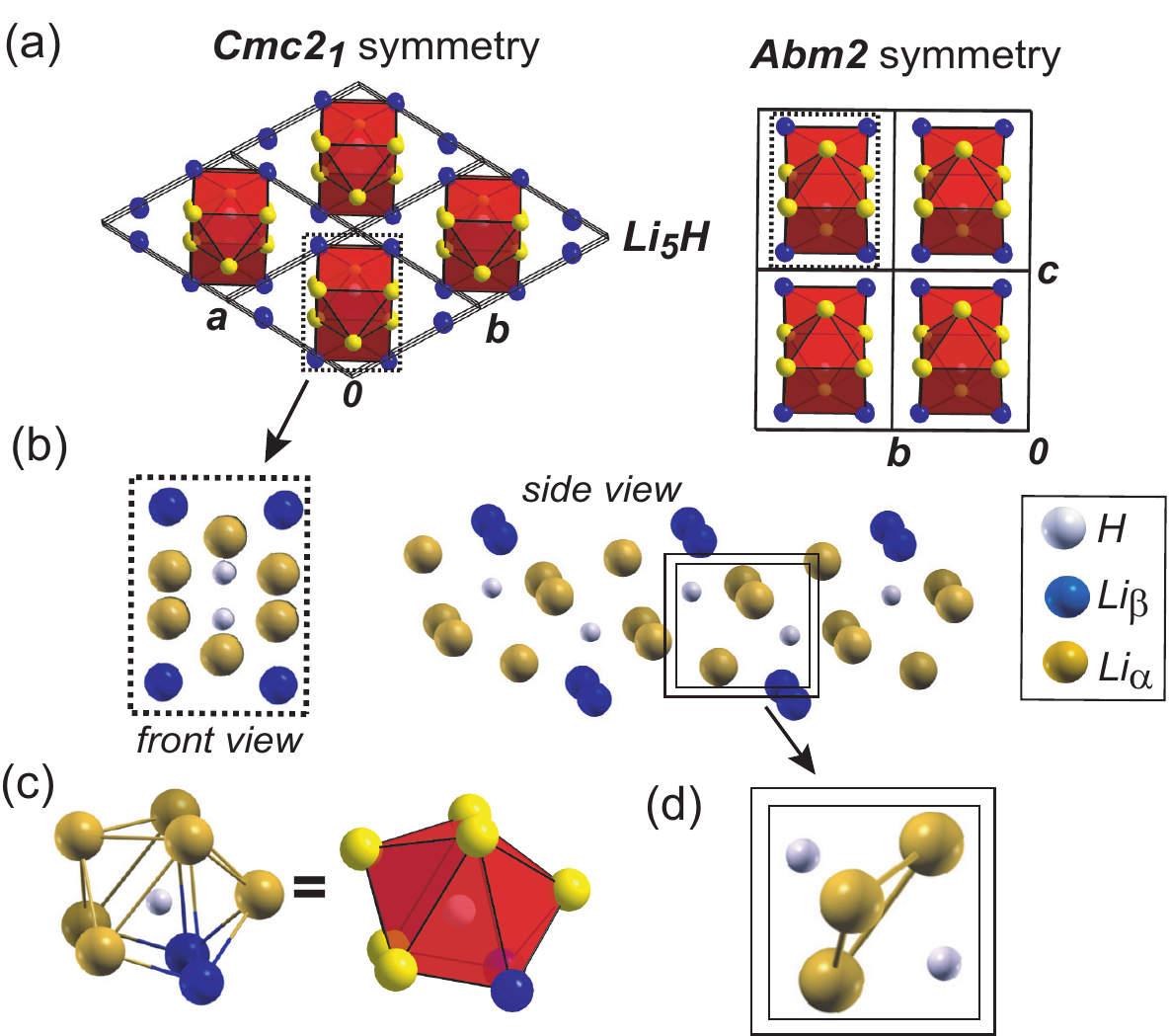}
\end{center}
\caption{(a) Extended structures \cite{footnote} of the two isoenthalpic Li$_5$H structures found between 50-100~GPa and (b) the chain--like building blocks common to both systems. These chains are made up of (c) Li$_8$H clusters which are fused at the faces as illustrated in (d) by the Li$_\alpha$ atoms. 
\label{fig:figure1}}
\end{figure}

Recently, we have shown that under pressure the alkali metal polyhydrides, MH$_n$ with $n>1$, become stable with respect to decomposition into MH and H$_2$  at 100, 25, and 2~GPa for lithium,\cite{Zurek:2009c} sodium,\cite{Zurek:2011d}, and rubidium, \cite{Zurek:2011h} respectively. LiH$_n$ is particularly interesting because the most stable structure above 150~GPa, LiH$_6$, was found to be metallic already at 1~atm.\cite{Zurek:2009c} 
Elemental lithium also exhibits unusual behavior: when compressed its melting temperature initially decreases so that between 40-60~GPa it has the lowest melting point among the elemental metals. \cite{Guillaume:2011a} These reasons prompted us to computationally explore the alkali--metal/hydrogen phase diagram further, specifically its lithium--rich region. Even though the structures we find, some of which are illustrated in Figure\ \ref{fig:figure1}, show no evidence for significant metallicity or low--temperature melting behavior, they hint that pressure may be used to synthesize a new class of compounds whose atomic and electronic structures are remarkably similar to the well--known rubidium and cesium suboxides.\cite{Simon:1997a, Simon:2010a} Because of the similarities described below, we refer to these phases as lithium subhydrides.

\begin{figure} 
\begin{center}
\includegraphics[width=0.9\columnwidth]{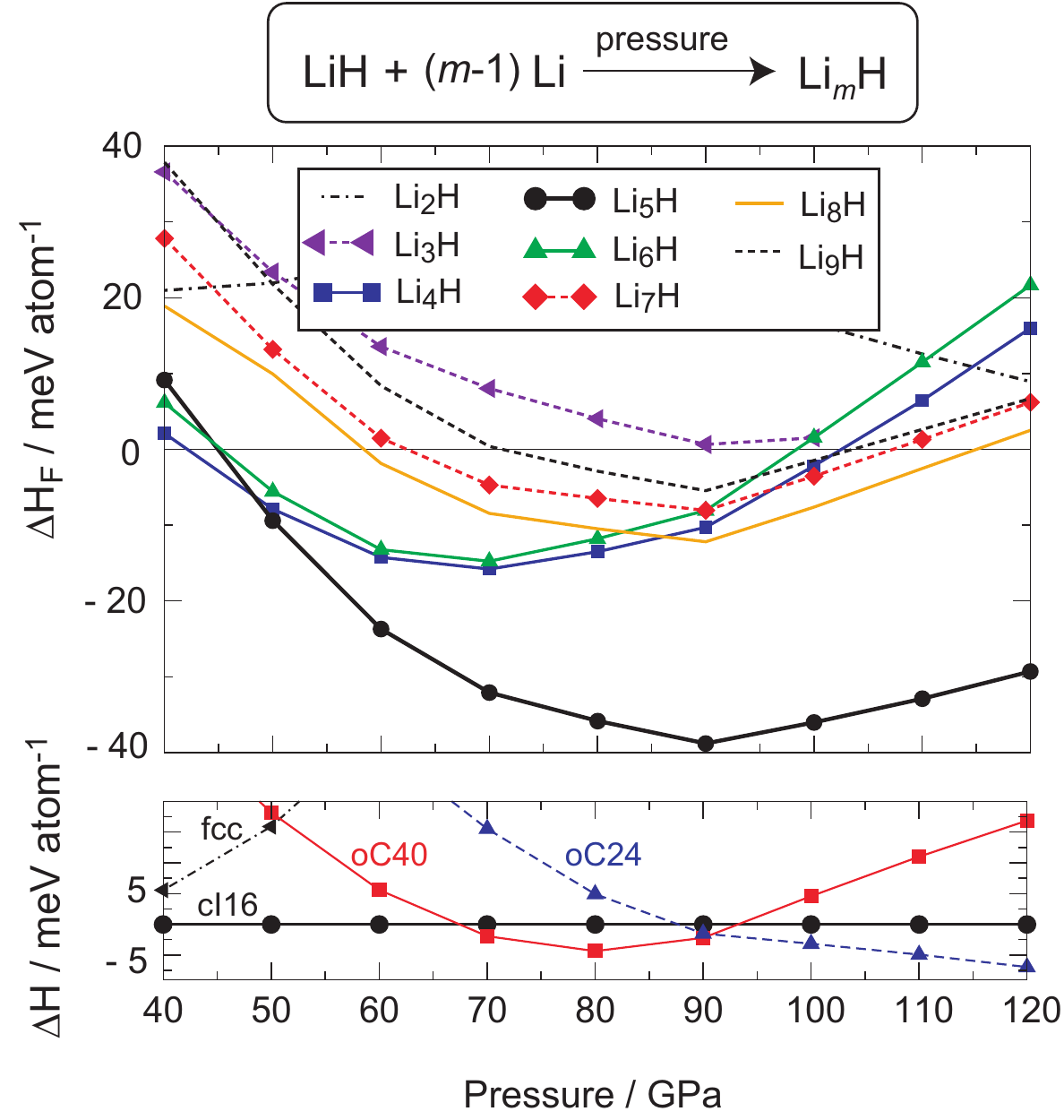}
\end{center}
\caption{$\Delta H_{F}$ for the reaction $\text{LiH} + ((m-1)\text{Li}) \rightarrow \text{Li}_m\text{H}$ versus pressure for various integer values of $m$. The enthalpy for Li is calculated for the most stable structures from Ref.\ \cite{Nelmes:2011a} and the enthalpy vs. pressure plot of the Li phases we considered is shown below. In this pressure range LiH is found in the NaCl structure.
\label{fig:figure2}}
\end{figure}

The open--source evolutionary algorithm (EA) \textit{XtalOpt} \cite{Zurek:2011a} has been employed to find low enthalpy structures of Li$_m$H with $2 < m < 9$ up to $P=$~100~GPa.\cite{comp} The calculated enthalpies of formation, $\Delta H_F$, of the most favorable structures for a given stoichiometry are provided in Figure\ \ref{fig:figure2}(a). The $\Delta H_F$ values are computed from the enthalpy differences between the right- and left- hand sides of the reaction shown at the top of Figure\ \ref{fig:figure2}; quantum and temperature effects are not included in this plot. In this pressure range (between 50-100~GPa) a number of Li$_m$H extended systems with $m=4-9$ are found to become (enthalpically) stable with respect to decomposition into Li and LiH, their would-be precursors in experiment. Moreover, the formation of LiH from lithium and molecular hydrogen is preferred up to at least 350 GPa  \cite{Zurek:2009c}. Our computations predict that two different Li$_5$H phases, one with $Cmc2_1$ (space group \#36) and one with $Abm2$ (space group \#39) symmetry (see Figure\ \ref{fig:figure1}(a)), are particularly stable in the sense that they have the most negative enthalpies of formation between 50--100~GPa, a pressure range which coincides with the metal--to--semiconductor transition \cite{Matsuoka:2009a} and the low melting point \cite{Guillaume:2011a} observed in pure lithium. Both structures have enthalpies within 3~meV/atom of each other within the entire pressure range studied. Phonon calculations confirm that both are mechanically stable within the harmonic approximation at 90~GPa and have the same zero--point energy. Beyond 180~GPa a Li$_2$H stoichiometry which is best described as a CsCl structure with missing hydrogen atoms has a lower $\Delta H_F$, but since this is also a region where lithium polyhydrides are stabilized,\cite{Zurek:2009c} we will focus our discussion on $m>3$.

Structurally, all of the Li$_m$H with $3<m<9$ are related to each other through common motifs: Li$_8$H cluster--like building blocks. The hydrogen atom lies near the center of these clusters and the lithium atoms make up the vertices of their enclosing polyhedra, the faces of one such polyhedron are colored red in Figure\ \ref{fig:figure1}(c). The most common cluster's geometric structure, which resembles a previously studied Li$_9$H cluster,\cite{footnote2} is shown in Figure\ \ref{fig:figure1}(c), as extracted directly from the Li$_5$H extended system. It is present in all of the structures except for $m=4$.\cite{footnote3} This $C_s$--symmetry polyhedron has 11 faces (10 triangular and 1 rectangular) and is best described as a distorted bicapped trigonal antiprism. Note that two opposing Li-Li distances on its ``rectangular'' base are distorted such that there is no C$_2$ axis in this cluster.  

\begin{figure} 
\begin{center}
\includegraphics[width=\columnwidth]{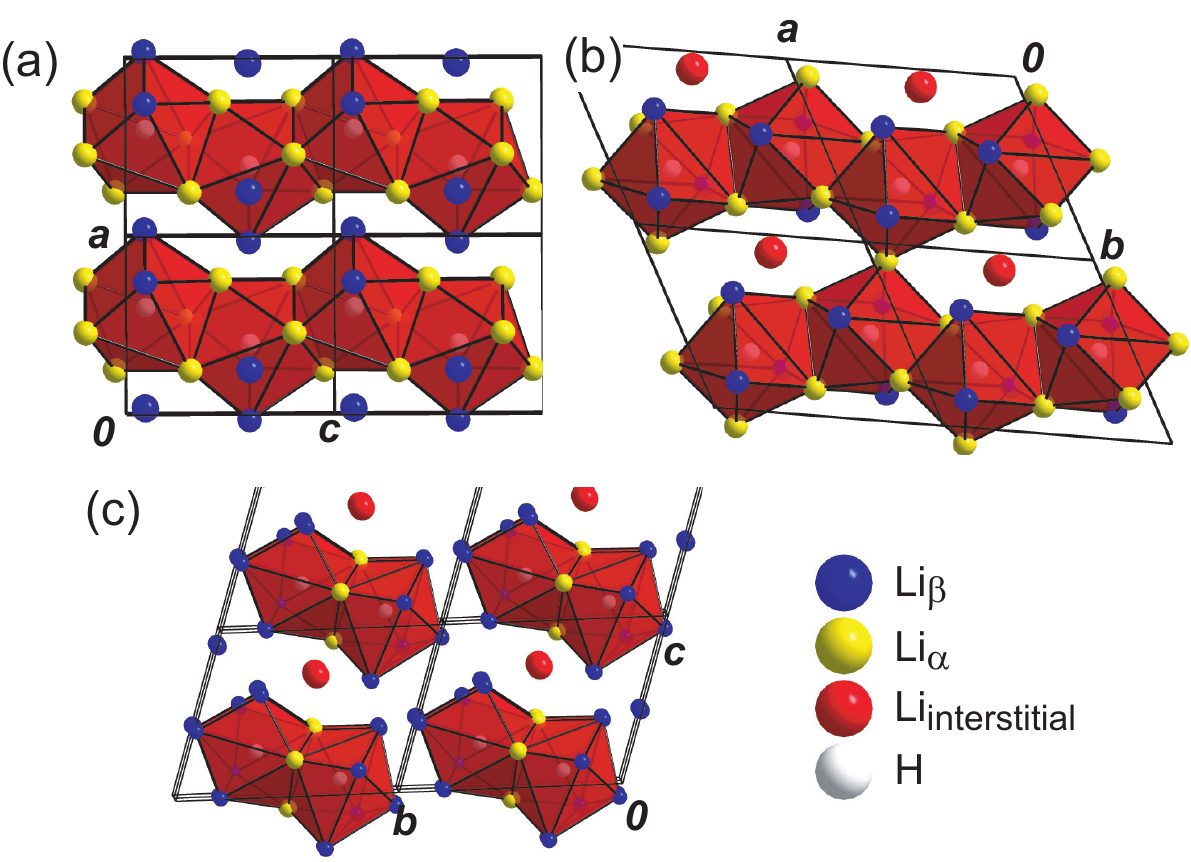}
\end{center}
\caption{ The underlying chain--like polyhedral building blocks of (a) $Cmc2_1$--Li$_5$H and (b) Li$_6$H at 90~GPa. The fused polyhedron pairs in Li$_7$H at 90~GPa are shown in (c).
\label{fig:figure3}}
\end{figure}

The Li$_8$H unit is also qualitatively similar to the most stable Li$_8$B isomer found in an extensive investigation of Li$_m$B clusters.\cite{Dixon:2011a} A [Li$_6$B]$^+$ cluster was found to be particularly stable relative to the rest and was noted to correspond to a closed electronic shell (magic number) within the phenomenological shell model,\cite{Cohen:1984a} as does [Li$_8$H]$^+$. We found this important because the building blocks of the suboxides, Rb$_9$O$_2$ and Cs$_{11}$O$_3$, have been compared to giant alkali metal (superalkali) atoms which may combine with excess metal atoms yielding a type of intermetallic compound.\cite{Simon:2010a} The neutral building block of the subhydrides, Li$_8$H, has the correct electron count ($8+1$) so that it can be thought of as a superalkali atom, a cluster with one more electron than a magic number (eight in this case). Moreover, interstitial lithium and/or hydrogen atoms (atoms which do not belong to a Li$_8$H building block) are observed in Li$_m$H with $m=(3, 6-9)$ (see Figs.\ \ref{fig:figure3}(b) and (c)). Thus, the extended systems predicted in this work have both structural and electronic similarities to the suboxides. Interestingly rare--earth halide cluster complexes almost always possess an endohedral atom as well.\cite{Zimmermann:2010a}

In Li$_5$H, the Li$_8$H clusters are fused at the faces in a chain--like arrangement, as illustrated in Figure\ \ref{fig:figure3}(a); the two Li$_5$H phases only differ in the relative stacking of the chains (see Figure\ \ref{fig:figure1}(a,b)). Li$_3$H and Li$_6$H are constructed in a similar fashion, except in Li$_6$H the polyhedra are fused via edges and vertices (see Figure\  \ref{fig:figure3}(b)). Similarly, Li$_4$H can also be thought of as a network of edge-- and vertex--sharing clusters. Li$_m$H with $m$=7--9  are made up of pairs of face--sharing polyhedra, as illustrated for Li$_7$H in Figure\ \ref{fig:figure3}(c). The range of Li$_m$H structures we considered (all of which are predicted to be thermodynamically unstable with respect to decomposition into Li$_5$H and Li/LiH) can all be represented as arrangemnts of Li$_8$H building blocks with lithium atoms possibly in the interstital regions.

Here lies the most striking similarity to the alkali metal suboxides: a structural one. The suboxides' basic building blocks, Rb$_9$O$_2$ and Cs$_{11}$O$_3$, are formed from two Rb$_6$O or three Cs$_6$O octahedra; they are similar to the fused polyhedron pairs shown in Figure\ \ref{fig:figure3}(c) and are the only constituents of the simplest suboxides. Different stoichiometries, such as Rb$_6$O=Rb$_9$O$_2 \cdot$Rb$_3$ and Cs$_7$O=Cs$_{11}$O$_3\cdot$Cs$_{10}$, are made up of the same basic units with excess Rb or Cs atoms interspersed between them, like Li$_7$H discussed above. The underlying octahedra can also be fused such that they form chains in more complex subnitrides or suboxoindates, such as NaBa$_3$N \cite{Simon:1997a} or Cs$_9$InO$_9$,\cite{Simon:2009a} just like Li$_5$H. 

\begin{figure}[h] 
\begin{center}
\includegraphics[width=\columnwidth]{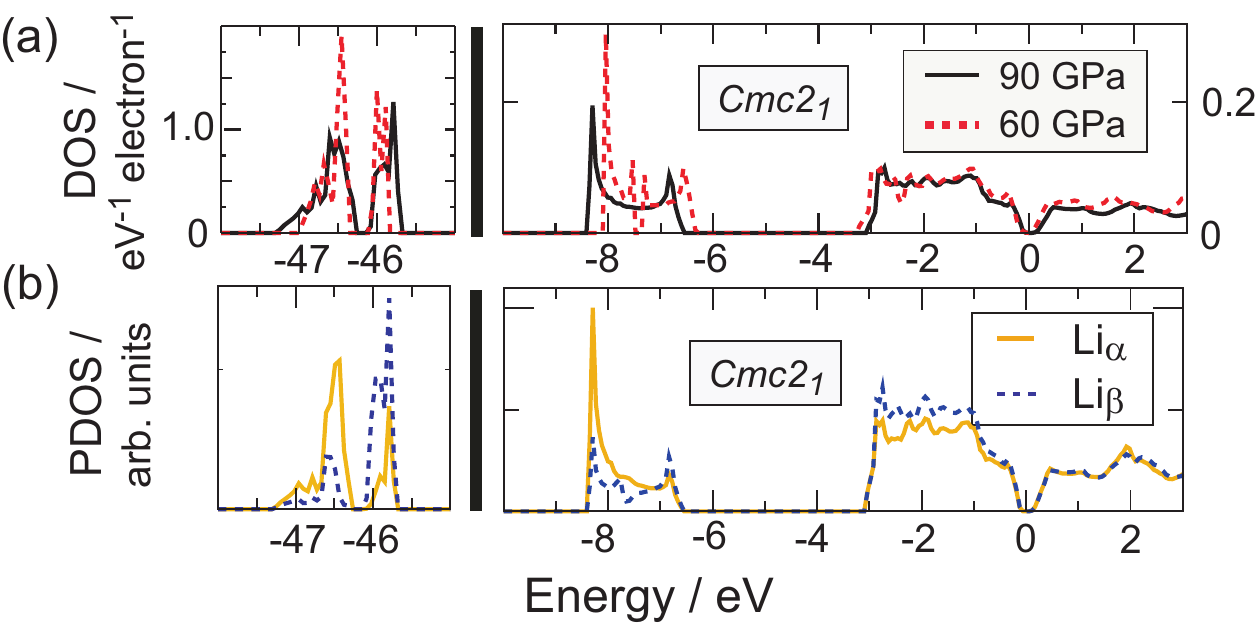}
\end{center}
\caption{(a) The total electronic densities of states (DOS) for $Cmc2_1$ Li$_5$H at 90~GPa (solid line) and 60~GPa (dashed line). The DOS plots for $Abm2$ Li$_5$H are qualitatively similar and are given in the SI. The left and right plots have different scales. (b) The site--projected densities of states (PDOS) onto the Li$_\alpha$ and Li$_\beta$ atoms in $Cmc2_1$--Li$_5$H at 90~GPa. The Fermi energy, $E_F$, is set to zero.
\label{fig:figure4}}
\end{figure}

In analogy to the suboxides and subnitrides, one can assume that because of the large electronegativity difference between lithium and hydrogen, the lithium subhydrides can be thought of as (Li$^+$)$_m$(H$^-$)$\cdot$(e$^-$)$_{m-1}$, exhibiting ionic bonding on the inside and metallic bonding on the outside of the fused Li$_8$H clusters yielding an overall metallic character. The calculated Bader charges of -0.96 on the hydrides and +0.68 on the lithium atoms agree with the ionic bonding, and the ubiquitous presence of the Li$_8$H building blocks suggests strong intra--cluster atomic interactions. The DOS of the two Li$_5$H phases at 60 and 90~GPa show these are semimetals or are weakly metallic, see Figure\ \ref{fig:figure4}(a). Furthermore, the existence of two iso--enthalpic Li$_5$H phases suggests that, like the suboxides, the bonding between the chains is weak.   

\begin{figure} [h!]
\begin{center}
\includegraphics[width=\columnwidth]{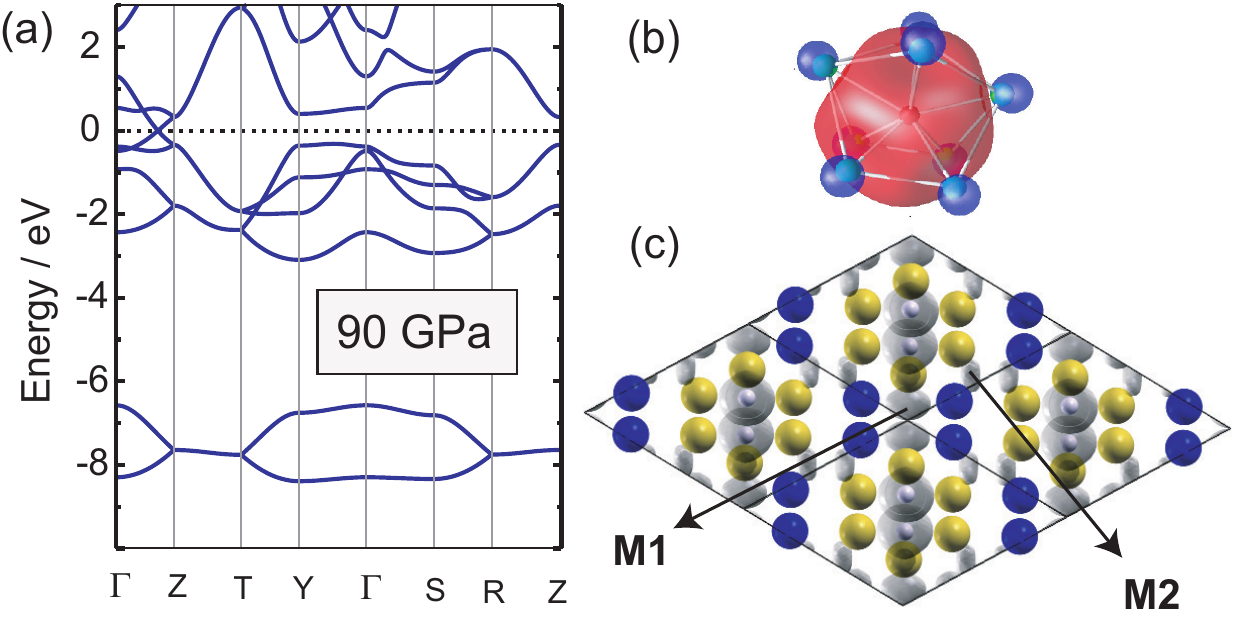}
\end{center}
\caption{(a) Band structure of $Cmc2_1$--Li$_5$H at 90~GPa, (b)  the Wannier Function which yields the two sets of  bands between -9 and -6 eV (isovalue$=\pm$~0.04~$a_B^{-3/2}$, where $a_B$ is the Bohr radius), and (c) an ELF=0.87 isosurface. In (c), the two characteristic maxima in the interstitial region of the lattice are labelled M1 and M2 and are shown in greater detail in the SI. 
\label{fig:figure5}}
\end{figure}

Under pressure the valence bands of Li$_5$H narrow slightly, the semi--core (sub--valence) bands do not change much, and the core Li $1s$ bands broaden, much like in elemental lithium. Also like lithium, the density of states near the Fermi energy, $E_F$, has mostly Li $p$ character. However, at 60~GPa Li$_5$H has a lower DOS at $E_F$ than Li--$cI16$ and at 90~GPa the subhydrides remain semimetallic but Li--$oC40$ is a semi-conductor. The band--structure of $Cmc2_1$--Li$_5$H in Figure\ \ref{fig:figure5}(a) shows that along the high--symmetry lines two conical--like bands, with a linear dispersion, cross $E_F$. In $Abm2$--Li$_5$H the band crossing occurs slightly above $E_F$, so this phase is somewhat more metallic.

The bands lying between -9 and -6 eV are primarily hydridic, but some lithium character is observed. The Li$_\alpha$ atoms, which belong to a face that is shared between two adjacent Li$_8$H clusters, contribute more than the Li$_\beta$ to the low energy portion of the sub--valence bands, as  shown in the PDOS in Figure\ \ref{fig:figure4}(b). In order to investigate this further, a Wannier function (WF) was constructed which describes these bands almost perfectly (see the SI) using the $N$MTO method.\cite{Zurek:2005a} This WF, shown in Figure\ \ref{fig:figure5}(b), consists of a sphere centered on the central H$^-$ with $p$--like tails on the surrounding lithium atoms, indicative of hybridization between the two. Interestingly, this WF is reminiscent of a $1s$ superatom orbital in a [Li$_8$H]$^+$ cluster with the same geometry as the building block used to construct the Li$_5$H phase (see the SI). It also shows that the bonding within the cluster is primarily ionic in nature.

We note that even though the WF which describes the sub--valence band of Li$_5$H in Figure\ \ref{fig:figure5}(b) is reminiscent of a superatom $s$--orbital, the valence DOS cannot easily be understood using such a picture. To extend the analogy, the remaining occupied bands would need to be described by superatom $p$--like states. Chain--like cluster arrangments of Li$_8$H, such as Li$_{18}$H$_3$, do qualitatively behave as such (see the SI), but we will show below that the comparison of the Li$_m$H phases with the suboxides (and high--pressure lithium) is perhaps a stronger one.

As expected, the electron localization function (ELF) is high around the hydridic atoms. There are, however, two characteristic pockets of increased electron density, labelled M1 and M2, with peak ELF values near 0.91 between the chains (see Figure\ \ref{fig:figure5}(c)) that can be attributed to the metallic bands. The ELF profile of the $Abm2$--Li$_5$H structure is similar and is consistent with the localization of valence electron density in the interstitial regions noted in pure compressed lithium.\cite{Ma:2011a,Neaton:1999a} The integrated charge densities within the M1/M2 basins (the charge density maxima identified from the Bader analysis) are 0.74/0.84 electrons in both Li$_5$H structures, notably smaller than the 1-2 electron occupancies in the recently proposed $C2cb$-40 structure for the pure Li--$oC40$ phase found experimentally at 90~GPa; this is suggestive that Li$_5$H is more metallic,\cite{Nelmes:2011a} as expected. However, the ELF between two M2 basins reaches values as high as 0.74, and the ELF between M1 and M2 does not exceed 0.68. The difference between these ELF values and the basins' peak ELF values of 0.91 may be suggestive of poor conductivity,\cite{Nelmes:2011a,Silvi:2000a} also as expected.    

The localization of metallic electrons in the areas between the chains are also characteristic of the suboxides.\cite{Simon:2010a} It is interesting to note that the M1 site shown in Figure\ \ref{fig:figure5}(c) is actually in the same local chemical environment as the hydrogen atoms. That is, the Li atoms around the site form the same exohedral structure as they do in the Li$_8$H cluster discussed above. Given that the 0.74 electrons assigned to the M1 site is comparable to the -0.96 charge assigned to the hydrides, this provides another trait (beyond the 1--dimensional behavior of the hydridic bands) which is unique to Li$_5$H among the Li$_m$H we considered: it allows the M1 sites, or ``pseudo-anions'', to be encapsulated in the same way as the hydrides. This also provides a simple physical link between the two Li$_5$H structures in the sense that they only differ in how the chains of hydrides and M1 sites are arranged about the lattice; in the $Cmc2_1$ structure they are distributed homogeneously while in $Abm2$ structure they are arranged into layers. The chemical difference between the Li$_\alpha$ and the  Li$_\beta$ atoms is that the former bridge two hydrides, whereas the latter bridge two M1 sites.

In conclusion, evolutionary algorithm searches coupled with first--principles computations predict that a new hydrogen--poor class of compounds, the lithium subhydrides (Li$_m$H with $m>1$) become stable with respect to decomposition into LiH and Li at $P\sim$~50~GPa. This stabilization occurs at pressures lower than those necessary to form their polyhydride counterparts.\cite{Zurek:2009c} All of the subhydrides are composed of superalkali, Li$_8$H, building blocks and exhibit ionic bonding on the inside of these clusters and metallic bonding in the interstitial regions between them. Analogies were drawn between the geometries and electronic structures of the phases predicted in this work and the well--known alkali metal suboxides.\cite{Simon:1997a, Simon:2010a} The most favorable stoichiometry, Li$_5$H, manifested in two iso--enthalpic phases which did not exhibit any imaginary phonon modes at 90~GPa. We find the most unique aspects of Li$_5$H relative to the other Li$_m$H structures we considered to be their thermodynamic and mechanical stability, the mixing of hydridic states with lithium states to create subvalence 1-D chain-like electronic behavior, and ordered Li cages around ``pseudo-anion''--like pockets of electron density resulting in a high symmetry Li sublattice relative to the other subhydride structures. Although we have not yet considered finite temperature effects nor the kinetic barriers required to break up their would-be precursors (Li or LiH), their enthalpic and mechanical stabilities suggest they may potentially be made in high--pressure experiments. Furthermore, these results hint that there may be other novel stoichiometries of subhydrides to be explored under pressure and, perhaps, other lithium-rich or suboxide materials.
  
\section*{Acknowledgements}
We acknowledge the NSF (DMR-1005413) for financial support, and the Center for Computational Research at SUNY Buffalo for computational support.

%\bibliography{pressure,vasp,hydrides,zurek,hions,lmto,nmto} %multicenter,RbH5_otherrefs,Pio}% Produces the bibliography via BibTeX.

%\bibliography{pressure,vasp,hydrides,zurek,hions,lmto,nmto,suboxides} 
%\bibliographystyle{angew}

\providecommand*\mcitethebibliography{\thebibliography}
\csname @ifundefined\endcsname{endmcitethebibliography}
  {\let\endmcitethebibliography\endthebibliography}{}

\clearpage

\noindent \textbf{Table of Contents Text:} \\ 
\textbf{Feeling the Pressure:} First principles calculations exploring the lithium--rich region of the lithium/hydrogen phase diagram show that the extended systems  Li$_m$H ($m=4-9$)  become stable at $\sim$50~GPa. These lithium subhydrides are constructed from superalkali Li$_8$H clusters, and are both structurally and electronically similar to  the well--known alkali--metal suboxides.  \\[2ex]

\noindent\textbf{Keywords:} hydrides, alkali metals, ab--initio calculations, high--pressure chemistry, bond theory \\[2ex]

\noindent \textbf{TOC graphic}
\begin{figure} [h!]
\begin{center}
\includegraphics[width=0.8\columnwidth]{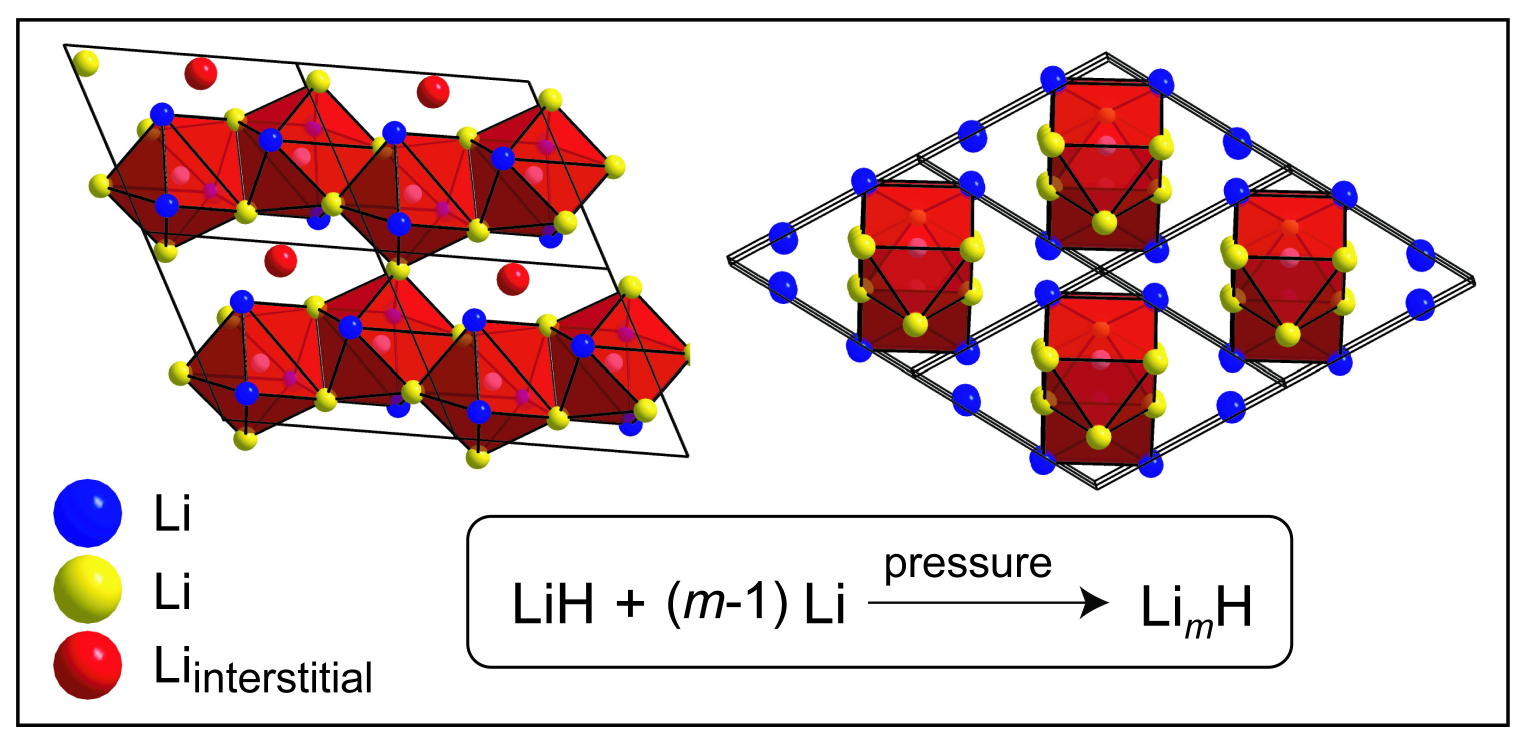}
\end{center}
\end{figure}

\end{document}